\documentclass[aip,rsi,reprint]{revtex4-1}

\usepackage{amsmath}

\usepackage{array}
\usepackage[pdftex]{graphicx}
\usepackage[normalem]{ulem}
\DeclareGraphicsExtensions{.pdf,.jpg}

\usepackage{makeidx}

\begin{document}

%\preprint{}

\title{Robust Mode Space Approach for Atomistic Modeling of Realistically Large Nanowire Transistors} %Title of paper



\author{Jun Z. Huang}
\email[]{junhuang1021@gmail.com}
%\thanks{}
\affiliation{Network for Computational Nanotechnology, Purdue University, West Lafayette, IN 47907, USA.}

\author{Hesameddin Ilatikhameneh}
%\email[]{}
%\thanks{}
\affiliation{Network for Computational Nanotechnology, Purdue University, West Lafayette, IN 47907, USA.}

\author{Michael Povolotskyi}
%\email[]{}
%\thanks{}
\affiliation{Birck Nanotechnology Center, Purdue University, West Lafayette, IN 47907, USA.}

\author{Gerhard Klimeck}
%\email[]{}
%\thanks{}
\affiliation{Network for Computational Nanotechnology, Purdue University, West Lafayette, IN 47907, USA.}


\date{\today}

\begin{abstract}
Atomistic quantum transport simulation of realistically large devices is computationally very demanding. The widely used mode space (MS) approach can significantly reduce the numerical cost but good MS basis is usually very hard to obtain for atomistic full-band models. In this work, a robust and parallel algorithm is developed to optimize the MS basis for atomistic nanowires. This enables tight binding non-equilibrium Green's function (NEGF) simulation of nanowire MOSFET with realistic cross section of $\rm 10nm\times10nm$ using a small computer cluster. This approach is applied to compare the performance of InGaAs and Si nanowire nMOSFETs with various channel lengths and cross sections. Simulation results with full-band accuracy indicate that InGaAs nanowire nMOSFETs have no drive current advantage over their Si counterparts for cross sections up to about $\rm 10nm\times10nm$.
\end{abstract}


\maketitle %\maketitle must follow title, authors, abstract and \pacs

\section{Introduction}
\begin{figure}[h]
\centering
\includegraphics[width=3.36in]{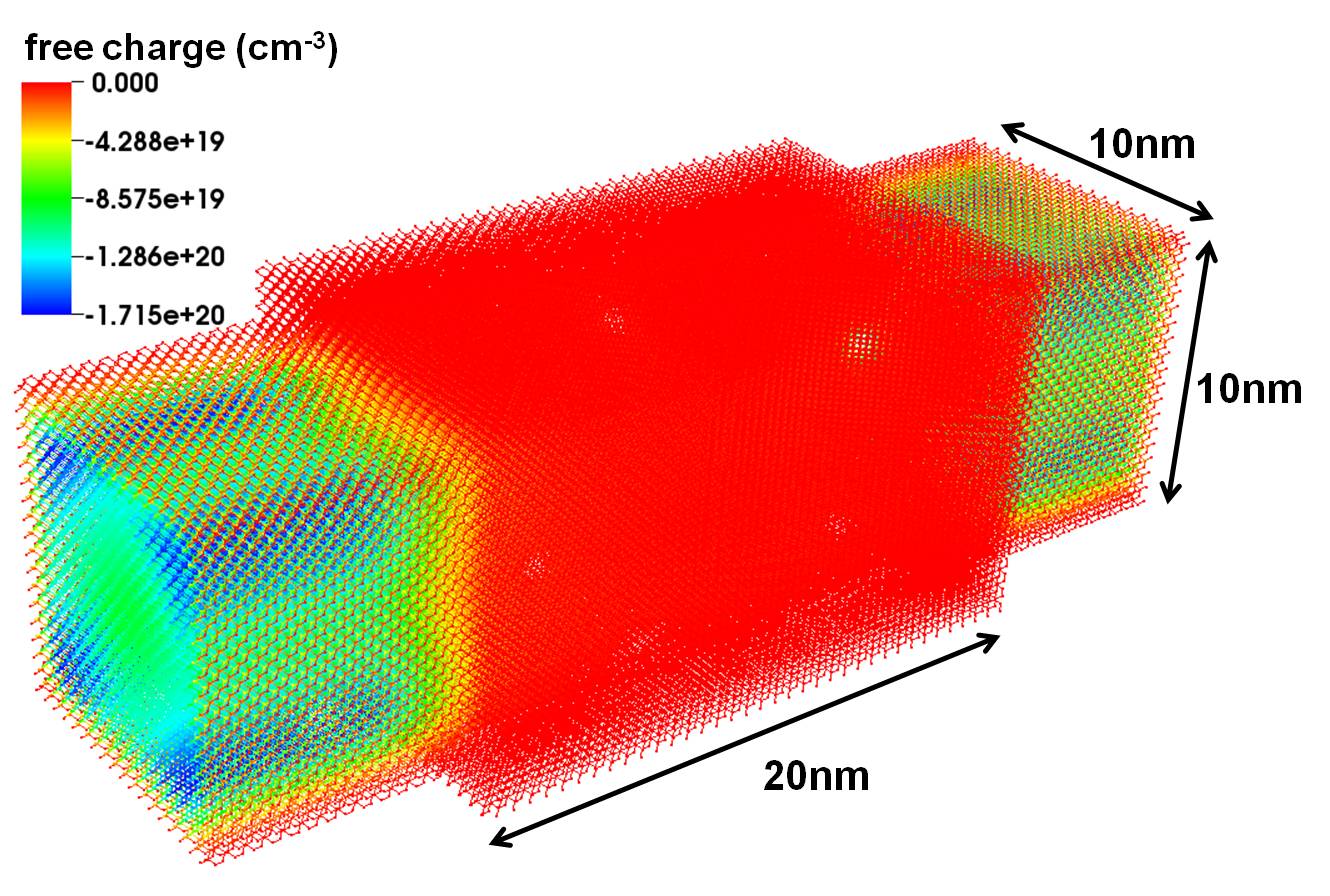}
\caption{An atomistic view of a gate-all-around Si nanowire nMOSFET with $10\rm{nm}\times10\rm{nm}$ channel cross section and $20\rm{nm}$ channel length. The Si nanowire (including the source and drain extensions) contains 165,888 Si atoms. The charge density distribution (at ON state) is obtained using the method developed in this study. The gate oxide is not included in the transport calculation but included in the Poisson equation.}
\label{atomistic_mosfet}
\end{figure}

As the key dimensions of nanotransistors continue shrinking to a few nanometers, quantum mechanical effects and atomistic details critically determine the device physics and performances. A well-established formalism to describe the quantum processes in open systems is the non-equilibrium Green’s function (NEGF) method~\cite{datta2005quantum,anantram2008modeling} or in the coherent limit the quantum transmitting boundary method (QTBM)~\cite{lent1990quantum,luisier2006atomistic}. The atomistic details can be captured if the system Hamiltonian is constructed with atomic resolution, using for instance the empirical tight binding (TB) models~\cite{tan2016transferable}. In fact, the TB-NEGF/QTBM solver has been the core functionality of the state-of-the-art nanoelectronics simulators~\cite{klimeck2010atomistic,steiger2011nemo5}. This solver, however, is computationally very demanding, becasue it requires (1) matrix inversions or solutions of generalized eigenvalue problems in the leads~\cite{huang2012methods}, and (2) matrix inversions or the solutions of linear systems in the device~\cite{luisier2006atomistic,lake1997single}. Unfortunately, the large number of atoms in realistically sized leads and devices multiplied with the large TB basis set results in huge Hamiltonian matrices. For example, a Si nanowire with $10\rm{nm}\times10\rm{nm}$ cross section and $35\rm{nm}$ length has 165,888 Si atoms (Fig.~\ref{atomistic_mosfet}); with $sp^3d^5s^*$ TB basis, {\it i.e.}, 10 orbitals per Si atom, the device Hamiltonian matrix size becomes 1,658,880. Even with the fast recursive Green’s function (RGF) algorithm~\cite{lake1997single} for the device matrix inversion, the numerical cost is still very expensive for large cross sections, because the computational time scales as $O\left(N_s^3N_l\right)$ and the memory consumption scales as $O\left(N_s^2N_l\right)$, where $N_s$ is the matrix dimension of a cross-sectional slab and $N_l$ is the number of slabs.

A widely used method to reduce the numerical cost is the mode space (MS) approach~\cite{polizzi2005subband}. In the MS approach, the matrix of each slab in the leads and in the device is transformed into incomplete MS which only consists of modes that are relevant to transport, significantly reducing the matrix dimensions. For effective mass models, the MS approach has been very successful~\cite{wang2004three,jin2006three}. For $k\cdot p$ models, the MS approach is also feasible through a more careful MS basis construction~\cite{shin2009full,huang2013model,huang2014model,huang2016quantum}. For TB models, however, the MS approach has been limited to special cases such as carbon nanotubes and graphene nanoribbons~\cite{guo2004toward,fiori2007coupled,grassi2011mode}, because in a general case it is very hard to obtain good MS basis. As noted by Mil'nikov {\it et al.}~\cite{mil2012equivalent}, the conventional way to contruct the MS basis usually leads to unphysical bands in the MS band diagram. Mil'nikov {\it et al.} further proposed a basis optimization process to remove the unphysical bands in a specified energy window~\cite{mil2012equivalent}. This method has led to a few successful applications including nanowire MOSFETs~\cite{mil2012equivalent} and tunnel FETs~\cite{afzalian2015mode}. It has also been extended to non-orthogonal Hamiltonian basis~\cite{shin2016density} and 2D ultra-thin-body (UTB) devices~\cite{jeong2016efficient}.

However, we found that the Mil'nikov method is not always stable, meaning that some unphysical bands may not be removed after the basis optimization process. Usually, one has to check the MS basis manually by looking at the MS band diagram, and if the MS basis is not good then the optimization process is restarted with a new set of input. This seriously restricts its practical applications for which reliable and automatic solution is sought. Another problem of the method is that the basis optimization involves matrix operations, whose numerical cost grows as $O\left(N_{uc}^3\right)$, where $N_{uc}$ is the Hamiltonian matrix size of a unit cell. As a consequence, the Mil'nikov method has only been applied to small nanowires with cross sections up to about $5.5\rm{nm}\times5.5\rm{nm}$~\cite{mil2012equivalent,afzalian2015mode,shin2016density}. In this paper, we substantially improve the Mil'nikov method, so that the basis optimization is always stable. Moreover, the optimization is parallelized using Message Passing Interface (MPI) and Open Multi-Processing (OpenMP). These allow us to obtain reliable TB MS basis of Si and III-V nanowires with cross sections larger than $10\rm{nm}\times10\rm{nm}$.

The obtained MS basis enables atomistic quantum transport simulation of large device structures. Here, we perform a systematic atomistic NEGF study of Si and $\rm{In}_{0.53}\rm{Ga}_{0.47}\rm{As}$ nanowire nMOSFETs with cross sections ranging from $4\rm{nm}\times4\rm{nm}$ to $10\rm{nm}\times10\rm{nm}$ and for channel lengths ranging from 4nm to 40nm. Nanowires allow gate-all-around (GAA) geometry which provides perfect gate electrostatic control over the channel and have been actively explored for future technology nodes. In addition, III-V materials, such as the $\rm{In}_{0.53}\rm{Ga}_{0.47}\rm{As}$, have recently drawn a lot of attention, for their potential of replacing Si as the channel materials of nMOSFETs. In fact, due to III-V materials' higher injection velocities and electron mobilities, they could deliver higher ON state current than Si at the same supply voltage~\cite{del2011nanometre}. Yet it is also known that III-V materials have lower density of states (DOS) which limits the charge density, and smaller effective mass which facilitates source-to-drain tunneling (SDT) at ultra-short channel lengths~\cite{mehrotra2013engineering,salmani2015design}. It is therefore very relevant to compare III-V and Si nanowire nMOSEFTs at various device sizes so as to provide useful information regarding the strength and weakness of different channel materials at different senarios. Due to the intensive numerical cost of atomistic quantum transport simulations, previous studies have only been limited to small nanowire cross sections~\cite{sylvia2012material,kim2015comprehensive}. For large cross sections, existing studies employed either top-of-the-barrier (TOB) model~\cite{wang2005performance,neophytou2008simulations}, or NEGF model in the simple effective mass approximation (EMA)~\cite{shin2007quantum,rau2016performance}. The atomistic TOB model~\cite{neophytou2008bandstructure} can capture atomistic effects and quantum confinement effects, but does not capture SDT, while NEGF in the EMA can capture SDT but does not capture atomistic and quantum confinement effects accurately.   

This paper is organized as follows. At first, the original Mil'nikov optimization method is briefly revisited in Section II. Then, the improved basis optimization scheme is detailed in Section III and validated in Section IV. In Section V, the method is applied to compare Si and $\rm{In}_{0.53}\rm{Ga}_{0.47}\rm{As}$ nanowire nMOSFETs with various sizes. Conclusions are drawn in Section VI.

\section{The Original Basis Optimization}
The essence of the original method of Mil'nikov {\it et al.}~\cite{mil2012equivalent} is summarized as follows. First, an initial MS basis set $\Phi$ is constructed by sampling and orthogonalizing the Bloch modes in the entire Brillouin zone. Sufficient modes should be sampled so that the initial MS band diagram contains all the physical bands in the energy window of interest. This is usually not a good basis set, since in its MS band diagram there are also unphysical bands. Then, a new basis state can be added into $\Phi$, which will alter the positions of the unphysical bands but will not affect the physical bands. The intention is that this new basis state can move some unphysical bands out of the energy window. This can be achieved by optimizing $C$, the expansion coefficients of the new basis state in a trial basis set $\Xi$, to minimize a cost function $\Delta F$~\cite{mil2012equivalent}:
\begin{eqnarray}\label{Delta_F}
\Delta F\left(C\right)=\frac{1}{2n_z}\sum_{i=1}^{n_q}\sum_{k=1}^{2n_z}\frac{C^TA\left(q_i,z_k\right)C}{C^TB\left(q_i,z_k\right)C}\left(z_k-\epsilon_c\right)\nonumber\\
+\left(C^TC-1\right)^2,
\end{eqnarray}
where
$q_i$ are the $n_q$ wave numbers in the 1D nanowire Brillouin zone, $\epsilon_c=\left(\epsilon_1+\epsilon_2\right)/2$ is the center of the energy window $\left[\epsilon_1,\epsilon_2\right]$, $z_k=\epsilon_c+\rho\exp(\frac{i\pi}{n_z}\left(k-\frac{1}{2}\right))$ are the $2n_z$ points in the complex $z$ plane along the circle with center $\epsilon_c$ and radius $\rho=\left(\epsilon_2-\epsilon_1\right)/2$. The cost function $\Delta F$ measures the change of number of states in the energy window after this new basis state is added into the initial basis set. Note that the number of states is calculated at a few representative wave numbers $q_i$ ($i=1,\cdots,n_q$). Therefore, minimizing this cost function would generate a new MS basis set that has less number of states (less bands), indicating that some unphysical bands must have been removed. 

The $A$ and $B$ matrices are, 
\begin{equation}\label{A}
A\left(q,z\right)=I_{M^\prime\times M^\prime}+\Xi^TH\left(q\right)\Phi\left[z-h\left(q\right)\right]^{-2}\Phi^TH\left(q\right)\Xi,
\end{equation}
\begin{eqnarray}\label{B}
B\left(q,z\right)=zI_{M^\prime\times M^\prime}-\Xi^TH\left(q\right)\Xi\\
-\Xi^TH\left(q\right)\Phi&\left[z-h\left(q\right)\right]^{-1}\Phi^TH\left(q\right)\Xi,\nonumber
\end{eqnarray}
where
\begin{equation}\label{hq}
h\left(q\right)=\Phi^TH\left(q\right)\Phi,
\end{equation}
and
\begin{equation}\label{Hq}
H\left(q\right)=H_0+We^{iq}+W^Te^{-iq},
\end{equation}
for which, the $H_0$ is the Hamiltonian of an isolated unit cell and the $W$ is the coupling Hamiltonian between two neighboring unit cells.

The $M^\prime$-dimensional trial basis set $\Xi$ is obtained by orthogonalizing the columns of the matrix
\begin{equation}\label{Xi}
\left[\left(1-\Phi\Phi^T\right)H\left(q=0\right)\Phi,\left(1-\Phi\Phi^T\right)H\left(q=\pi\right)\Phi\right],
\end{equation}
where $\left(1-\Phi\Phi^T\right)$ acts as a projector to the orthogonal complement to the space of $\Phi$. This trial basis constructed with only $q=0$ and $q=\pi$ offers enough degrees of freedom for the optimization.

\begin{figure}[h]
\centering
\includegraphics[width=3.36in]{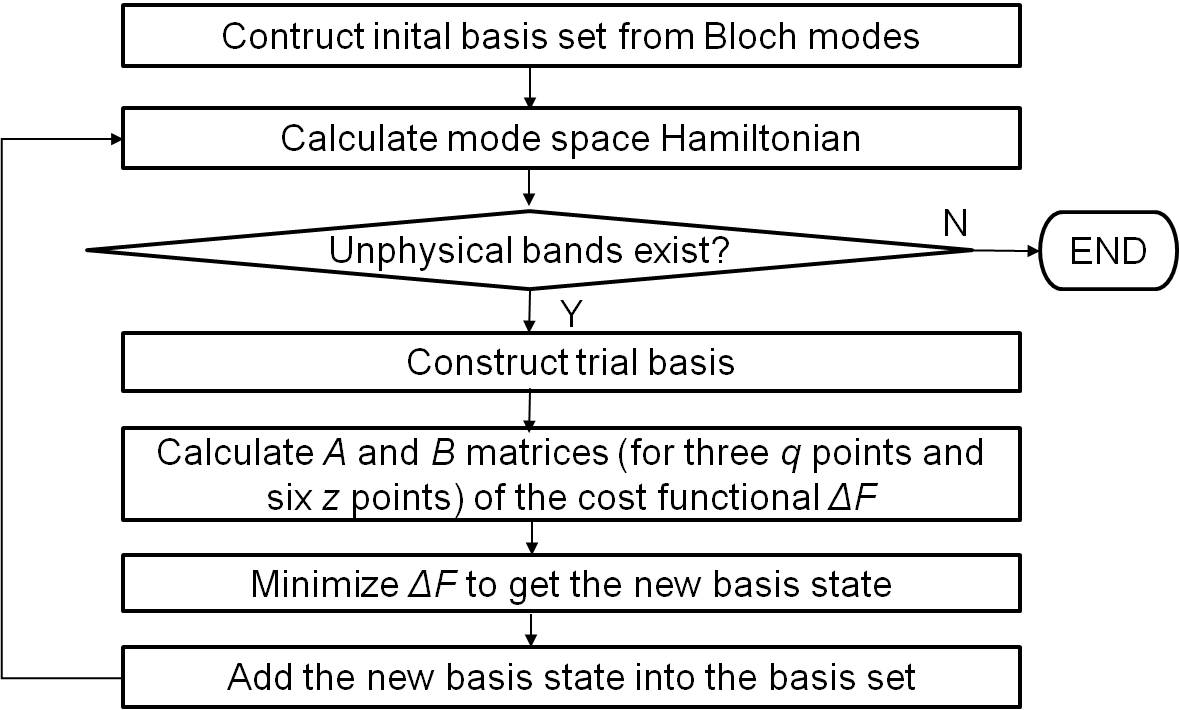}
\caption{A typical basis optimization flow of Mil'nikov {\it et al.}~\cite{mil2012equivalent}.}
\label{original_opt_flow}
\end{figure}

The above optimization can be repeated until all unphysical bands are moved out of the energy window. A typical basis optimization flow is shown in Fig.~\ref{original_opt_flow}, where $n_q=3$ and $n_z=3$ as suggested in the original paper~\cite{mil2012equivalent}.

Three issues arise in this optimization flow. First, how to check if an unphysical band exists? Simply comparing the MS and TB band structures at a few $q$ points is not sufficient, since unphysical bands can show up anywhere in a band diagram. Second, the optimization is performed for certain fixed representative $q$ points, there is no way to guarantee unphysical energies at other $q$ points are also removed. This leads to a serious problem: after a series of optimizations there are still unphysical bands appearing between the representative $q$ points. It turns out that the success of the optimization depends on many factors, such as the initial sampling of Bloch modes, the orthogonalization, the selection of representative $q$ points, {\it etc.} Third, the optimization is numerically very expensive due to the matrix multiplications and inversions, as evident from Eq.~\eqref{Delta_F} to Eq.~\eqref{B}.

\section{The Improved Basis Optimization}
To overcome the above three difficulties in the basis optimization, we have made the following improvements.

\begin{figure}[h]
\centering
\includegraphics[width=3in]{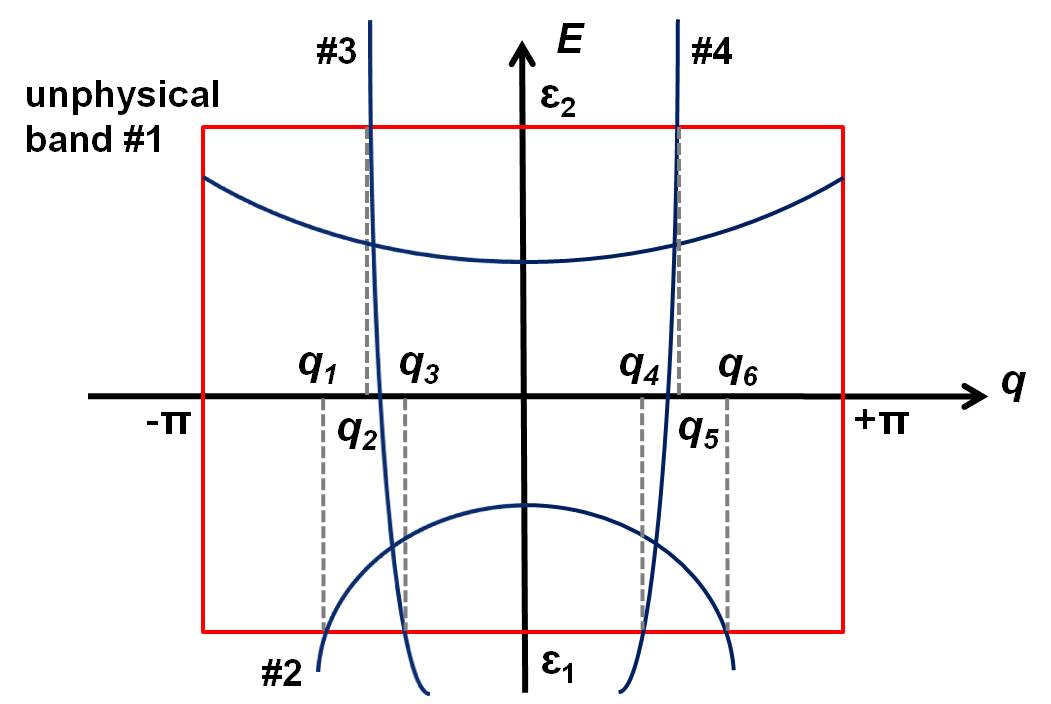}
\caption{Illustration of four possible unphysical bands. Any unphysical band must cross at least one of the four red lines.}
\label{unphysical_bands}
\end{figure}

First, we designed a reliable and efficient scheme to check the existence and identify the locations of the unphysical bands. The portion of the band diagram we are interested in is bounded by four lines: the Brillouin zone boundaries $q=+\pi$ and $q=-\pi$, and the energy window boundaries $E=\epsilon_1$ and $E=\epsilon_2$. It is observed that, any unphysical band that pollutes this portion of the band diagram, if exists, must pass through at least one of these four lines (Fig.~\ref{unphysical_bands}). Therefore, it is \emph{sufficient} to compare MS and TB band diagrams \emph{only} on these four lines. Specifically, we first solve and store the eigen-energies $E$ at $q=+\pi$ and $q=-\pi$, as well as the propagating (real) $q$ at $E=\epsilon_1$ and $E=\epsilon_2$, all in the TB space. Next, we do the same in the MS. Subsequently, these two sets of values are compared. If there is an extra $E$ at $q=+\pi$ or $q=-\pi$ of the MS band diagram, it means there is an unphysical band passing through $q=+\pi$ or $q=-\pi$. Similarly, if there is an extra $q$ at $E=\epsilon_1$ or $E=\epsilon_2$ in the MS band diagram, it means there is an unphysical band passing through that $q$. If these two set of values are identical, the MS band diagram will be free of any unphysical band in the energy window. In the case of Fig.~\ref{unphysical_bands}, the unphysical bands pass through $q=-\pi,q_1,\cdots,q_6,+\pi$.

\begin{figure}[h]
\centering
\includegraphics[width=3.36in]{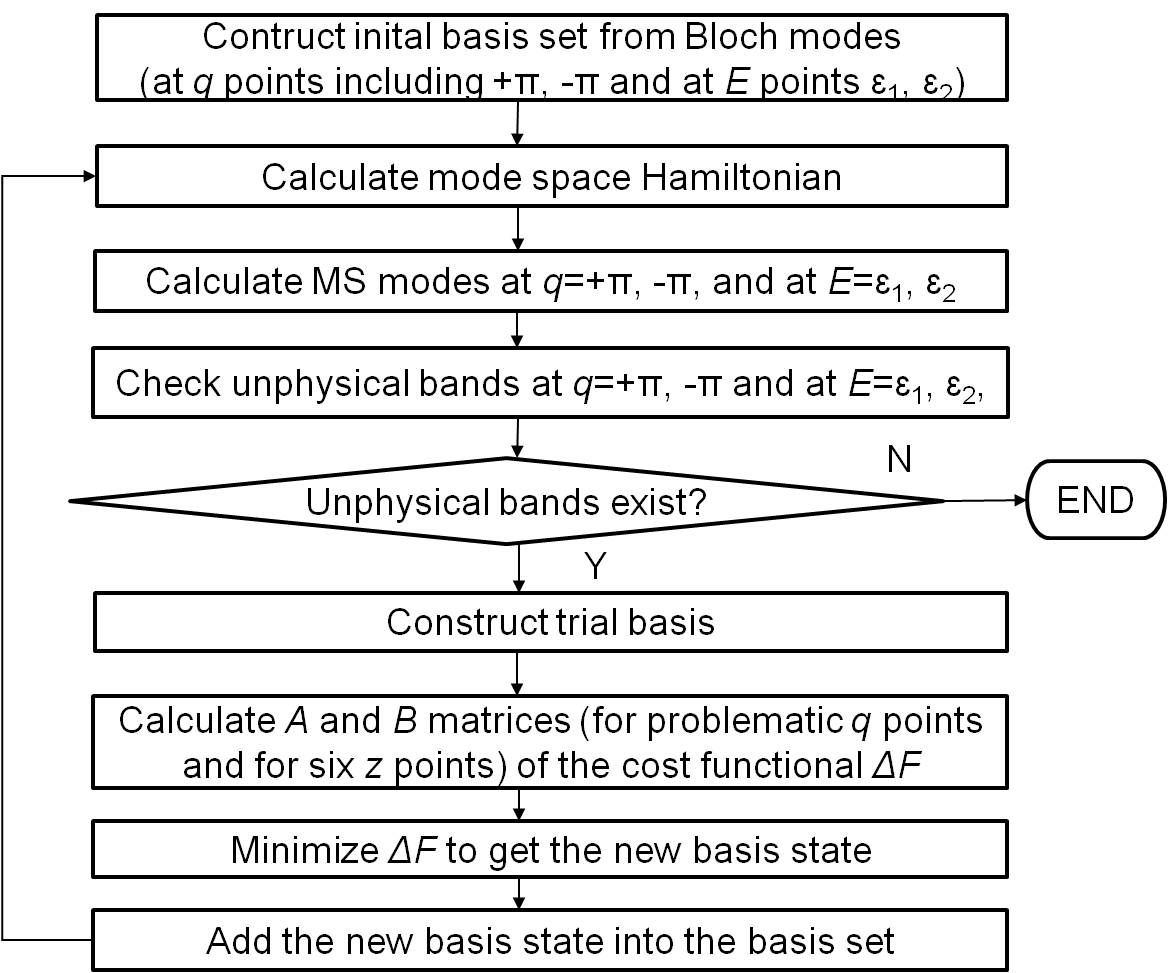}
\caption{The improved basis optimization flow, which guarantees the final MS basis is a good basis.}
\label{improved_opt_flow}
\end{figure}

Second, to break the limitations of fixed optimization $q$ points, we allow change of $q$ points of Eq.~\eqref{Delta_F} in each optimization. Specifically, the $q$ points identified by the above comparison will be the optimization $q$ points. In other words, the optimization always targets at the problematic $q$ points where there are unphysical energies. In the case of Fig.~\ref{unphysical_bands}, the optimization $q$ points will be $q=-\pi,q_1,\cdots,q_6,+\pi$. Note that the number of optimization $q$ points identified, {\it i.e.}, $n_q$, can be very large at beginning thus we have restricted $n_q$ in each optimization to control the cost. After a few optimizations, the $n_q$ becomes smaller, and eventually $n_q=0$ meaning that all the unphysical bands are safely removed. The improved optimization flow is shown in Fig.~\ref{improved_opt_flow}.

Third, the optimization efficiency is improved by parallelization. Two of the steps in Fig.~\ref{improved_opt_flow} are identified as the numerical bottlenecks: (1) the initial basis construction, which involves normal eigenvalue problems at different $q$ and generalized eigenvalue problems at different $E$; and (2) the calculation of $A$ and $B$ matrices, which involves expensive matrix operations and needs to be done for different $q$ and $z$ points. Two levels of parallelization are thus implemented: first, different $q$ and $E$, or $q$ and $z$ points are distributed to different MPI processes; second, each matrix inversion and multiplication are parallelized with openMP threads using the Intel(R) Math Kernel Library. Note that the generalized eigenvalue problems for solving the propagating modes at a given $E$ can be sped up by the shift-and-invert strategy with proper selection of the shift targets~\cite{huang2013model}.

\section{Method Validation}
The method is implemented in the NEMO5 software~\cite{steiger2011nemo5} and validated in two steps. The first step is to show that the improved basis optimization can generate good MS basis for large nanowires; and second, the MS-NEGF simulation using the optimized basis can generate accurate I-V curves for transistors with significant speed up.
\begin{figure}[h]
\centering
\includegraphics[width=3.36in]{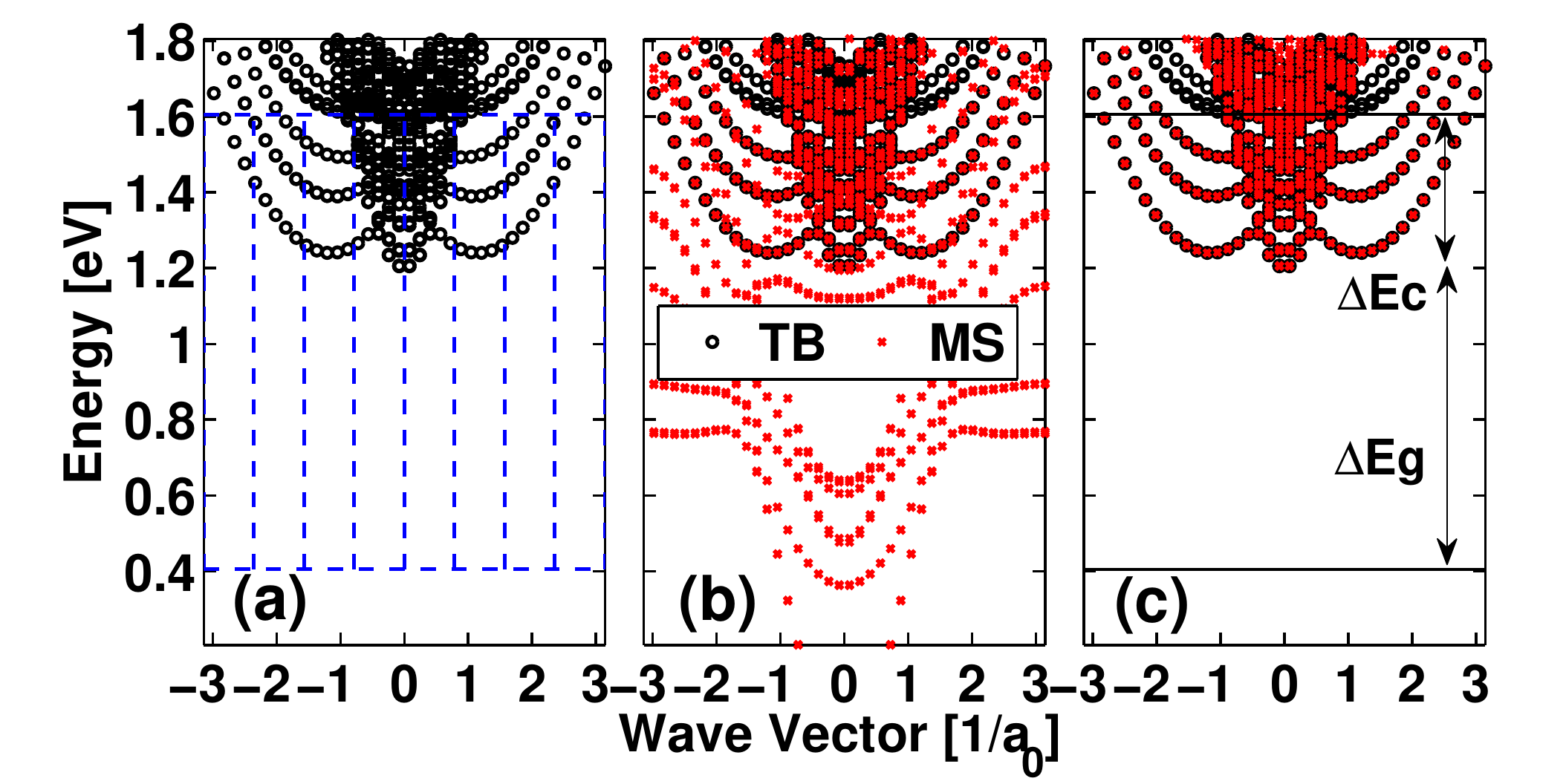}
\caption{Basis optimization of the $5.43\rm{nm}\times5.43\rm{nm}$ Si nanowire. (a) Sampling lines for the Bloch modes. (b) MS band diagram using the initial MS basis. (c) MS band diagram using the optimized MS basis.}
\label{Ek_sampling_opt}
\end{figure}

\subsection{Basis Optimization}
A [100]-oriented Si nanowire with $5.43\rm{nm}\times5.43\rm{nm}$ cross section (10 unit cells in each of the cross-sectional direction which is [010] or [001]) is considered here. The TB basis is $sp^3d^5s^*$ without spin-orbit coupling. Spin-orbit coupling has a minor effect on the conduction band thus is neglected. For simplicity, the electrical potential in the nanowire is set to be zero everywhere. The energy window to be optimized is chosen to be $\left(\Delta E_g,\Delta E_c\right)=\left(0.8\rm{eV},0.4\rm{eV}\right)$, where $\Delta E_g$ and $\Delta E_c$ are the energy ranges below and above the confined conduction band edge. As shown in Fig.~\ref{Ek_sampling_opt} (a), the Bloch modes are sampled at nine $k$ lines evenly distributed in the entire Brillouin zone and at two $E$ lines which are the boundaries of the energy window. To form an initial MS basis, the sampled Bloch modes are orthogonalized with linearly dependent modes removed (through a singular value decomposition and the columns with small singular values are discarded). The MS band diagam using this initial MS basis (158-dimensional) is shown in Fig.~\ref{Ek_sampling_opt} (b). As expected, there are many unphysical bands which are not present in the TB band diagram. The optimized MS basis (224-dimensional, after adding 66 basis states) leads to MS band diagram as plotted in Fig.~\ref{Ek_sampling_opt} (c), where the unphysical bands are clearly removed. The error between the MS and TB band diagrams in Fig.~\ref{Ek_sampling_opt} (c) is further quantified to be less than 0.1meV/mode within the optimization window.

\begin{figure}[h]
\centering
\includegraphics[width=3.36in]{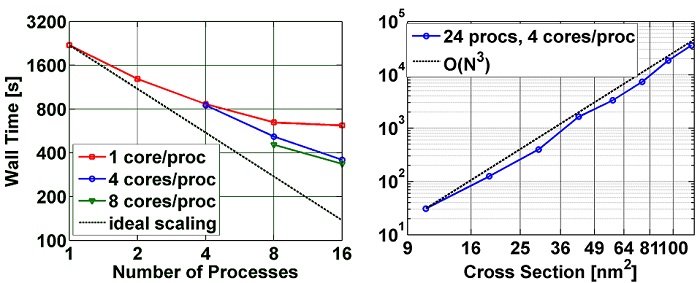}
\caption{Wall time of the basis optimization versus number of MPI processes (using 1, 4, and 8 cores per MPI process) for the $5.43\rm{nm}\times5.43\rm{nm}$ Si nanowire (left), and versus Si nanowire cross section (using 24 MPI processes and 4 cores per MPI process) (right).}
\label{Ek_opt_scaling}
\end{figure}

The scaling of the basis optimization time versus number of cores and versus cross-sectional sizes is plotted in Fig.~\ref{Ek_opt_scaling}. It is seen that the MPI parallelization, together with the multithreading, reduces the wall time by up to 7 times. For larger cross sections, the parallelization efficiency is expected to be better. It is also observed that the optimization time scales with cross section roughly following the $O\left(N^3\right)$ rule.

\begin{table*}
\centering
\caption{Basis optimization and Poisson-NEGF results of the $4.34\rm{nm}\times4.34\rm{nm}$ Si nanowire MOSFET.}
\label{tab:negf_sim_details} 
\begin{tabular}{p{0.8cm}|p{1.8cm}p{1.5cm}|p{1.4cm}p{1.9cm}p{2.7cm}|p{1.4cm}p{1.9cm}p{2.7cm}}
\hline
        &Basis Size &Opt. Time & Iterations (OFF) & Total Time (OFF) & Current (OFF)  & Iterations (ON) & Total Time (ON) &  Current (ON) \\
\hline
TB  &5120 &  & 6 & 129371s                      & 5.4814pA                 & 7 & 151140s                           & 14.912$\mu\rm{A}$\\
MS1 &147 (2.87\%) &93.8s & 6 & 262s ($494\times$)      & 5.4580pA (0.43\%)   & 8 & 339s ($446\times$)           & 14.759$\mu\rm{A}$ (1.03\%)\\
MS2 &191 (3.73\%) &156s & 6 & 343s ($377\times$)      & 5.4739pA (0.14\%)   &8  & 464s ($326\times$)           & 14.781$\mu\rm{A}$ (0.88\%)\\
\hline
\end{tabular}
\end{table*}

\subsection{MS-NEGF Simulation}
The self-consistent Poisson-MS-NEGF simulation flow for a nanowire transistor is briefly summarized as follows. First, the MS basis for a nanowire slab (with zero potential) is optimized as described above. Then, the device Hamiltonian matrix (including potential from the Poisson solver) is transformed into the MS, assuming the same MS basis for all the device slabs. Next, the MS-NEGF equations are solved with the RGF algorithm. Here, the RGF calculations for different energy points are distributed to different MPI processes. Finally, the charge density matrix obtained in the MS is transformed back into the real space (only diagonal elements are computed) for the Poisson solver.

Here, we consider a gate-all-around (GAA) Si nanowire nMOSFET with $4.34\rm{nm}\times4.34\rm{nm}$ cross section, 8.68nm gate length, $1\times10^{20}\rm{cm}^{-3}$ souce and drain doping density, and 1nm equivalent oxide thickness (EOT). Such a small device allows us to obtain TB-NEGF data for benchmarking our MS-NEGF models. To further save the computational resources required for TB-NEGF simulation, we only benchmark transistor OFF state ($V_{\rm{DS}}=0.5\rm{V}$ and $V_{\rm{GS}}=-0.25\rm{V}$) and ON state ($V_{\rm{DS}}=0.5\rm{V}$ and $V_{\rm{GS}}=0.45\rm{V}$) instead of the full I-V curve. To show the dependence of the accuracy of the MS-NEGF simulation on the optimization energy window, the MS basis is optimized for two energy windows, $\left(\Delta E_g,\Delta E_c\right)=\left(0.8\rm{eV},0.4\rm{eV}\right)$ and $\left(0.8\rm{eV},0.5\rm{eV}\right)$, denoted by MS1 and MS2.

The numerical results are summarized in Table~\ref{tab:negf_sim_details}. The calculations (basis optimization and NEGF simulation) are performed on 6 nodes with 4 MPI processes per node and 4 cores per MPI process. Each node is comprised of dual 8-core Intel Xeon-E5 CPUs. As the energy window increases, the MS basis size of a unit cell ($N_{\rm{MS}}$) increases and the basis reduction ratio ($N_{\rm{TB}}/N_{\rm{MS}}$) drops.  The basis optimization time and Poisson-MS-NEGF simulation time both increase with the energy window expansion. The accuracy of the drain current is improved if the speed up factor is reduced. For both MS cases, speed up factor of over $300\times$ and drain current error of less than 1\% are achieved. The basis optimization time is not included in calculaton of the speed up factor, since once the basis is obtained it can be stored and re-used for all bias points. It is also observed that the speed up factor is less than the estimated one from the complexity analysis, {\it i.e.}, $\left(N_{\rm{TB}}/N_{\rm{MS}}\right)^3/16$, because in the MS-NEGF algorithm there is an overhead due to the matrix transformations. The factor 16 is present because the MS RGF algorithm is based on unit cells, while the TB RGF can be done on atomic planes due to the nearest neighbor interaction. Note, that each unit cell in the [100] orientation has four atomic planes.

\begin{figure}[h]
\centering
\includegraphics[width=1.68in]{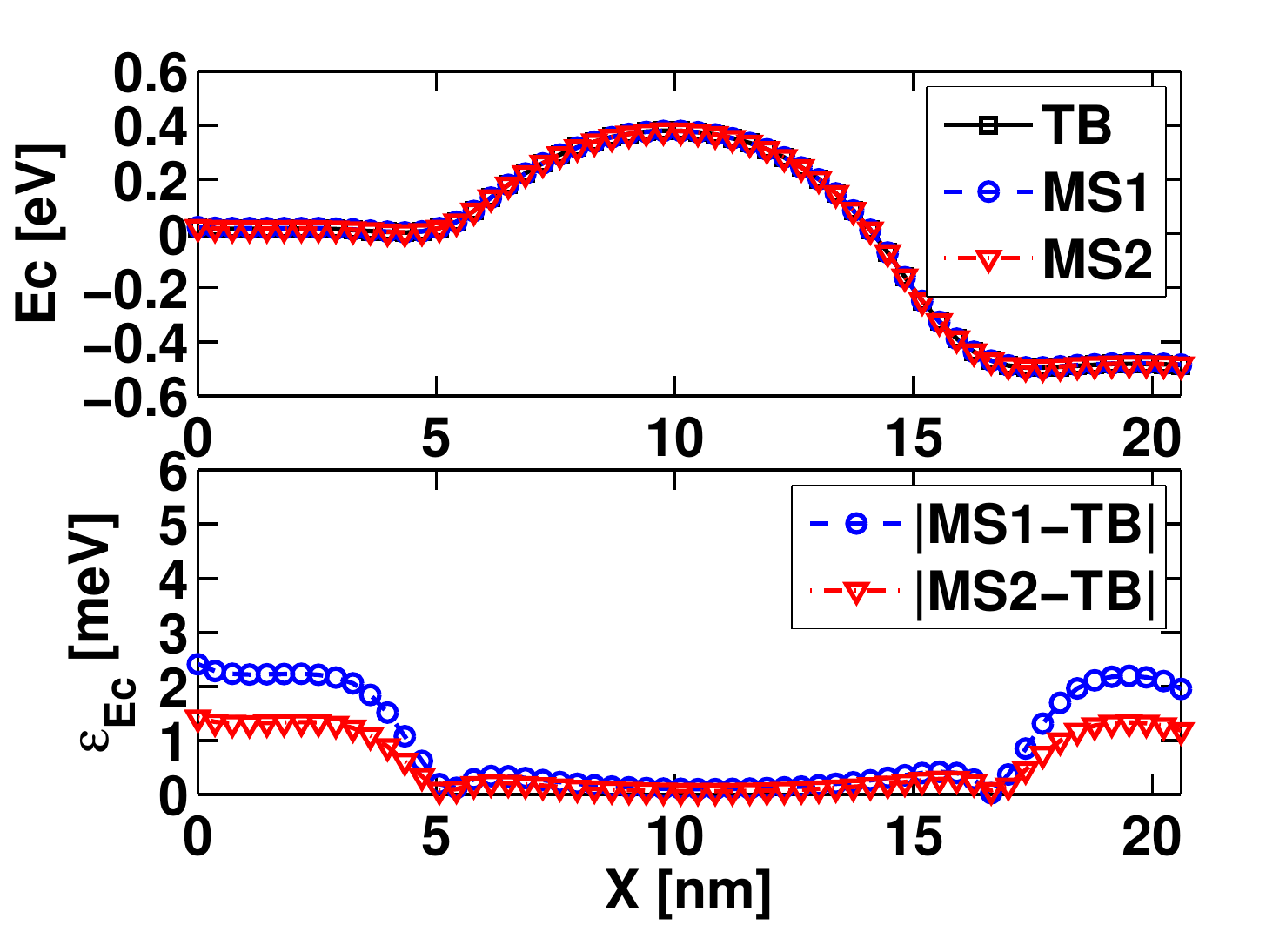}
\includegraphics[width=1.68in]{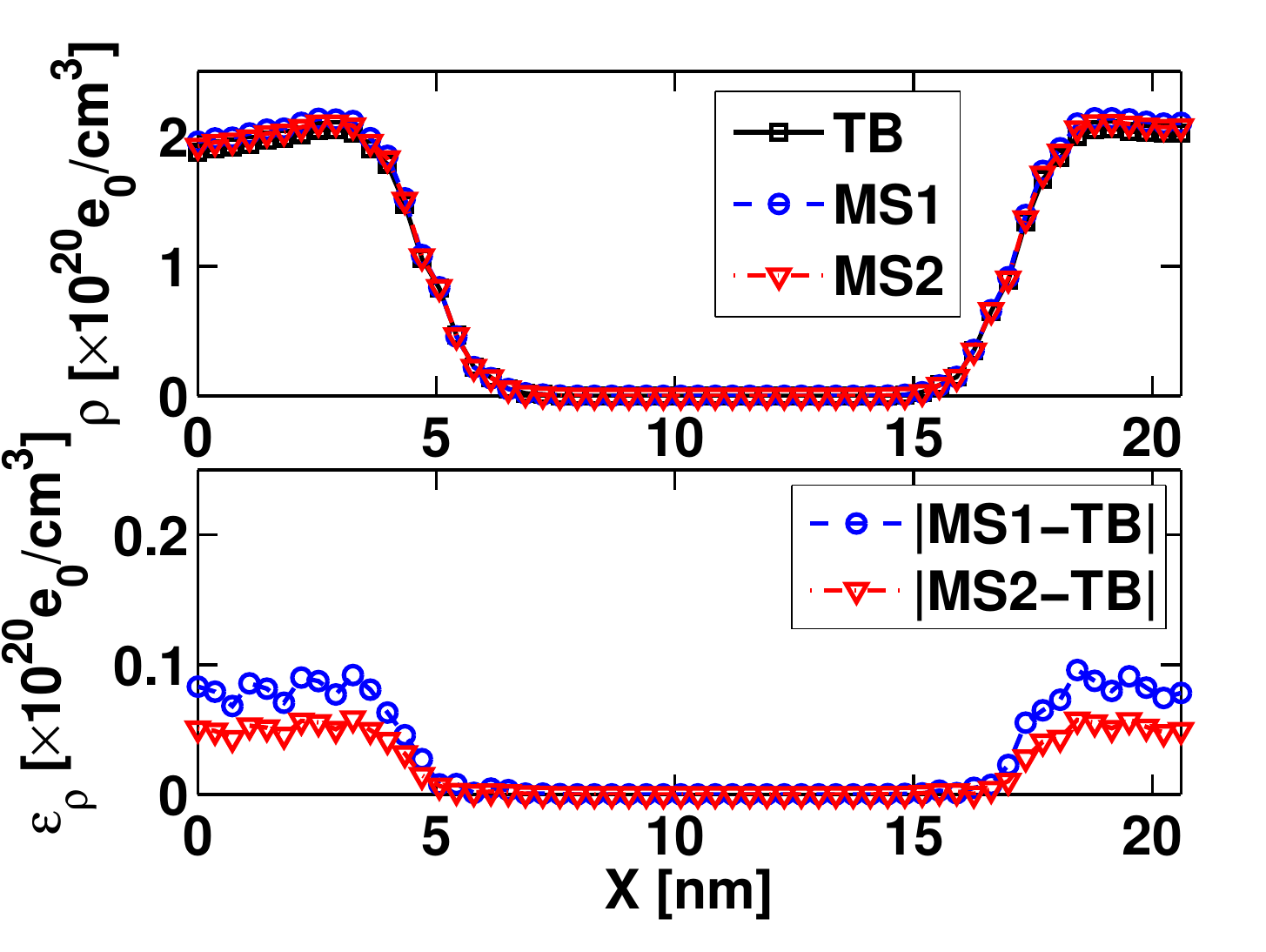}
\caption{Top left: potential, and top right: charge density, along the center of the nanowire transistor, obtained from TB, MS1, and MS2. Bottom left: absolute error of the potential, and bottom right: absolute error of the charge density, of MS1 and MS2, with respect to TB. This is for the OFF state.}
\label{error_OFF}
\end{figure}

\begin{figure}[h]
\centering
\includegraphics[width=1.68in]{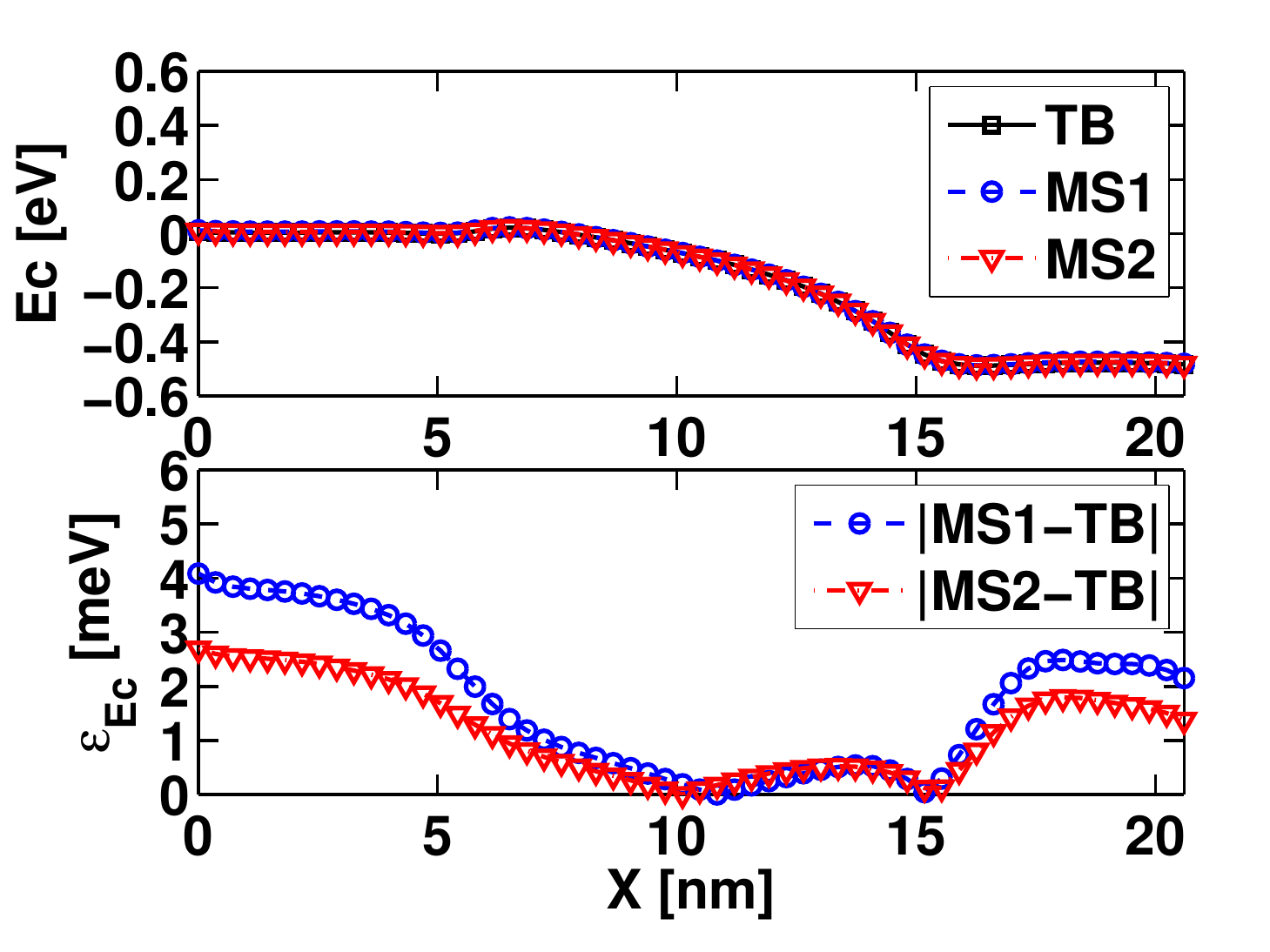}
\includegraphics[width=1.68in]{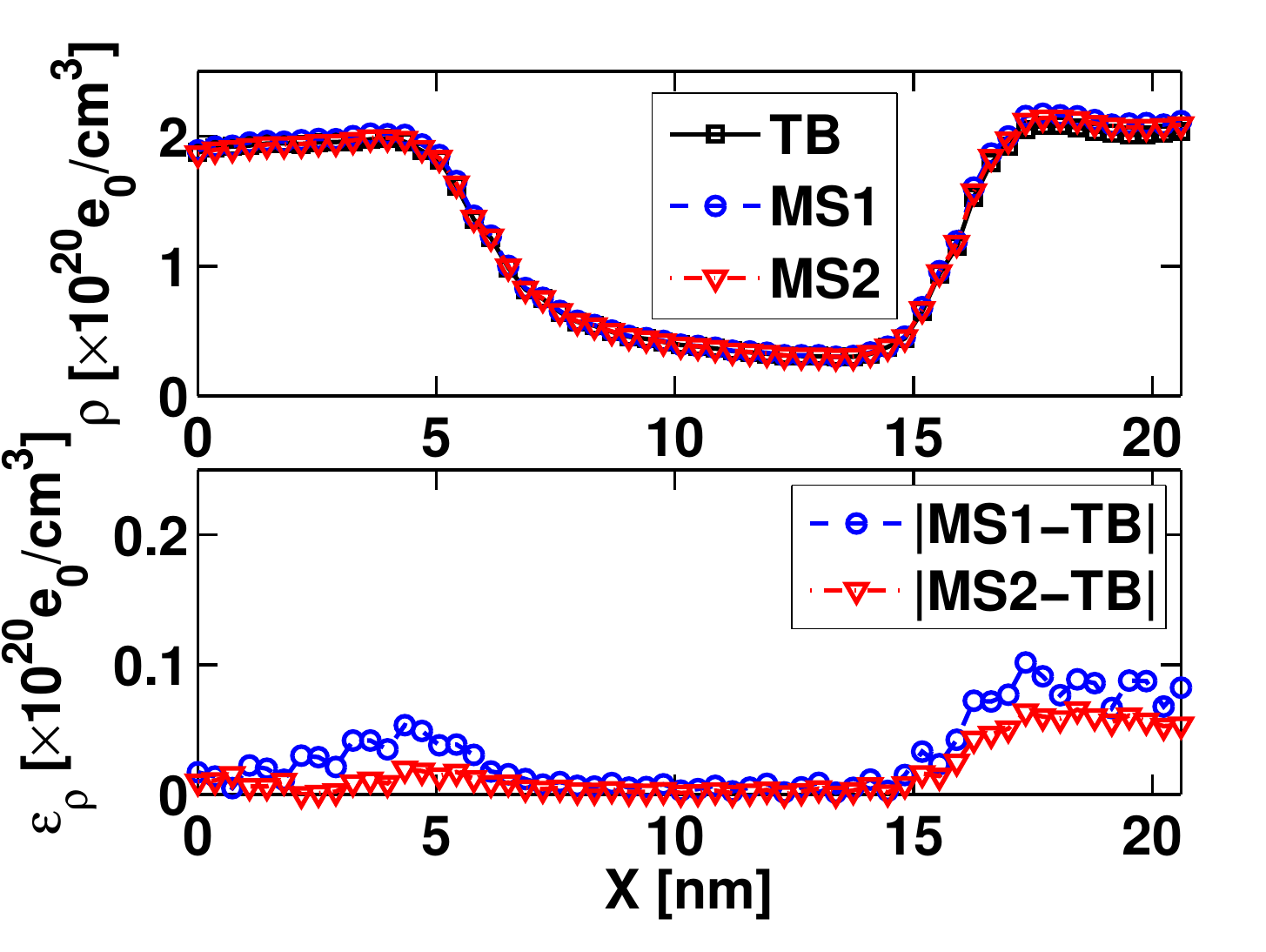}
\caption{The same as Fig.~\ref{error_OFF}, but for the ON state.}
\label{error_ON}
\end{figure}

The potential and charge density distributions are further plotted in Fig.~\ref{error_OFF} for the OFF state and in Fig.~\ref{error_ON} for the ON state. It can be seen that as the energy window (and number of modes) increases, the accuracy of potential and charge density from the Poisson-MS-NEGF solver also improves. The error is larger in the doped source and drain regions than in the channel region. Overall, the error of the potential is within a few meV, and the relative error of the charge density is within a few percent.

\section{Application: Si and InGaAs Nanowire Transistors}
\begin{figure}[]
\centering
\includegraphics[width=3.36in]{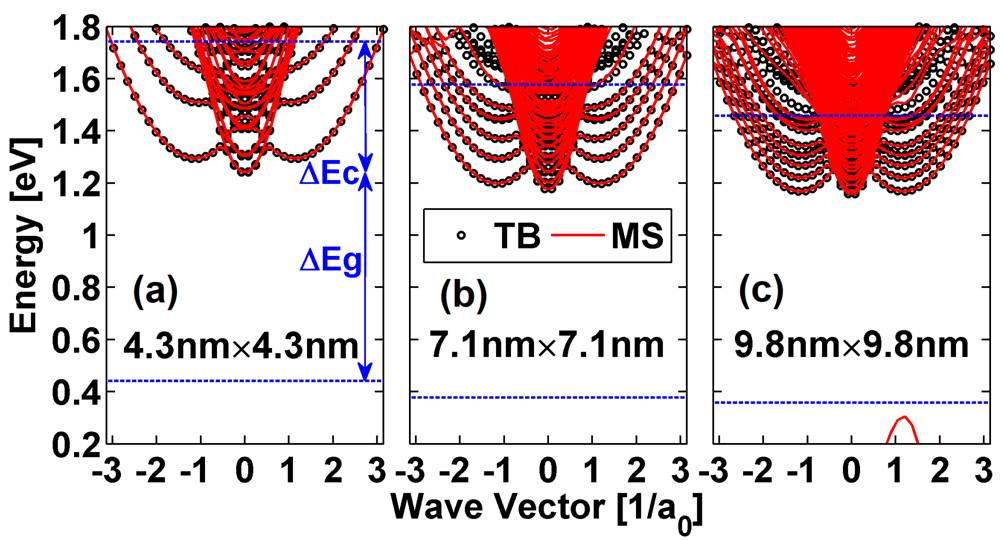}
\caption{Basis optimization for Si nanowires with three cross sections. (a) cross section $4.3\rm{nm}\times4.3\rm{nm}$, MS optimization window $\left(\Delta E_c,\Delta E_g\right)=\left(0.5\rm{eV},0.8\rm{eV}\right)$, basis reduction $N_{MS}/N_{TB}=191/5120=3.7\%$; (b) $7.1\rm{nm}\times7.1\rm{nm}$, $\left(\Delta E_c,\Delta E_g\right)=\left(0.4\rm{eV},0.8\rm{eV}\right)$, $N_{MS}/N_{TB}=374/13520=2.8\%$; and (c) $9.8\rm{nm}\times9.8\rm{nm}$, $\left(\Delta E_c,\Delta E_g\right)=\left(0.3\rm{eV},0.8\rm{eV}\right)$, $N_{MS}/N_{TB}=429/25920=1.7\%$.}
\label{Ek_cross_sections}
\end{figure}

\begin{figure}[]
\centering
\includegraphics[width=3.36in]{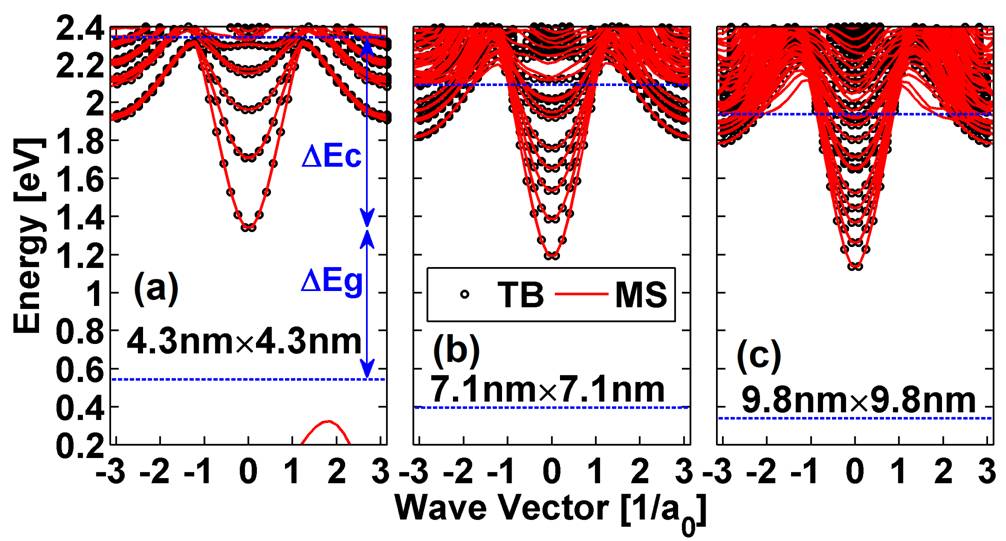}
\caption{Basis optimization for $\rm{In}_{0.53}\rm{Ga}_{0.47}\rm{As}$ nanowires with three cross sections. (a) cross section $4.3\rm{nm}\times4.3\rm{nm}$, MS optimization window $\left(\Delta E_c,\Delta E_g\right)=\left(1.0\rm{eV},0.8\rm{eV}\right)$, basis reduction $N_{MS}/N_{TB}=251/4500=5.6\%$; (b) $7.1\rm{nm}\times7.1\rm{nm}$, $\left(\Delta E_c,\Delta E_g\right)=\left(0.9\rm{eV},0.8\rm{eV}\right)$, $N_{MS}/N_{TB}=329/12010=2.7\%$; and (c) $9.8\rm{nm}\times9.8\rm{nm}$, $\left(\Delta E_c,\Delta E_g\right)=\left(0.8\rm{eV},0.8\rm{eV}\right)$, $N_{MS}/N_{TB}=343/22450=1.5\%$.}
\label{InGaAs_Ek_cross_sections}
\end{figure}

Three cross-sectional sizes are selected in this study: $4.3\rm{nm}\times4.3\rm{nm}$, $7.1\rm{nm}\times7.1\rm{nm}$, and $9.8\rm{nm}\times9.8\rm{nm}$, covering small, medium, and large cross sections. All the nanowires are oriented in the [100] direction and confined in the [010] and [001] directions. The TB basis is again $sp^3d^5s^*$ without spin-orbit coupling. The virtual crystal approximation is employed for the InGaAs alloy with its TB parameters linearly interpolated from its binary compounds. First, the MS basis sets are optimized, with the MS and TB band diagrams compared in Fig.~\ref{Ek_cross_sections} for the Si cases and in Fig.~\ref{InGaAs_Ek_cross_sections} for the $\rm{In}_{0.53}\rm{Ga}_{0.47}\rm{As}$ cases. Good matches are observed for all the cases, indicating that the basis optimizations are successful. Note that the basis reduction ratio is well controlled to under a few percents.

\begin{figure*}[]
\centering
\includegraphics[width=7in]{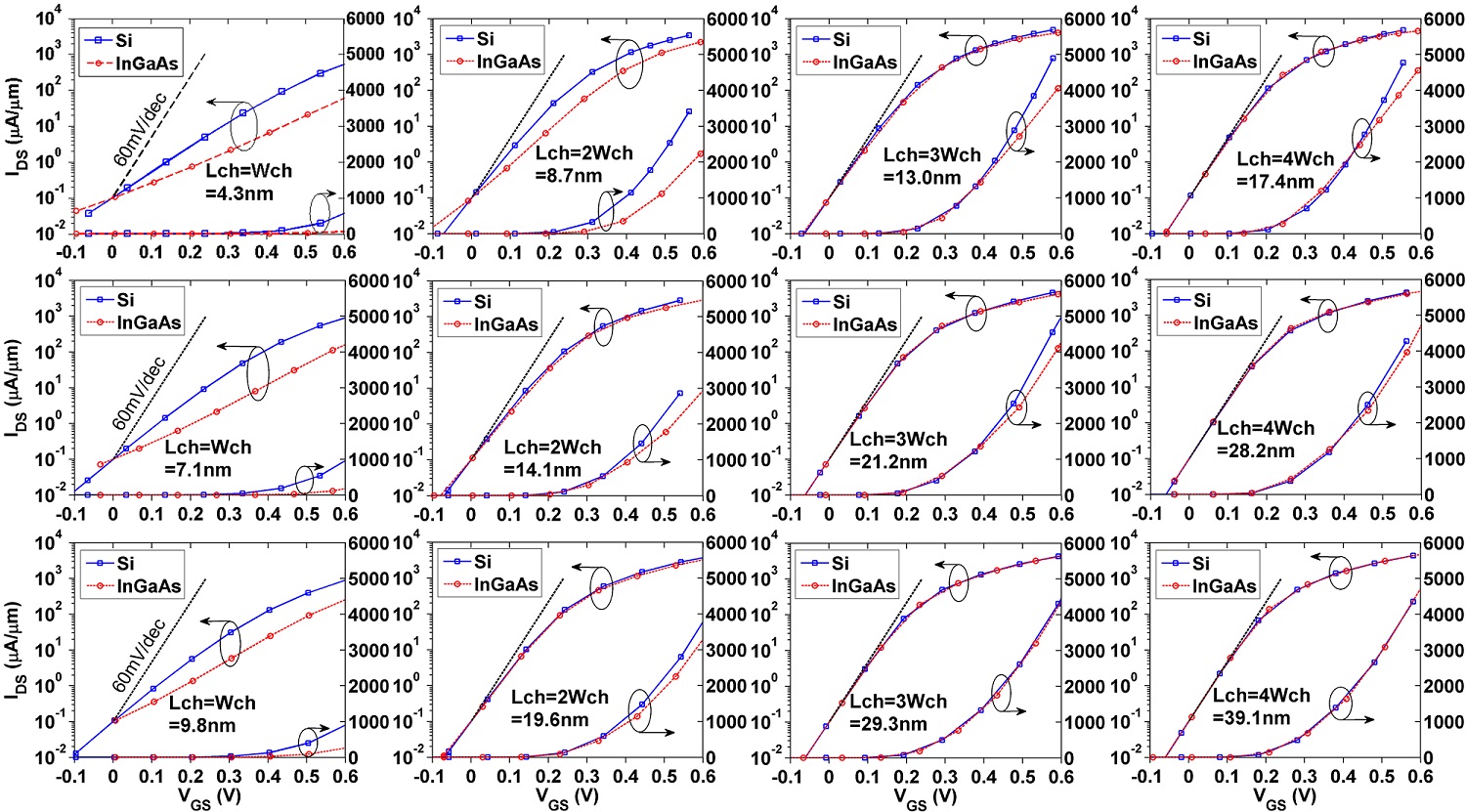}
\caption{$I_{\rm{DS}}-V_{\rm{GS}}$ ($V_{\rm{DS}}=0.5\rm{V}$) curves of the Si and InGaAs nanowire nMOSFETs, with threshold voltage adjusted to have $I_{\rm{OFF}}=0.1\mu\rm{A}/\mu\rm{m}$ at $V_{\rm{GS}}=0\rm{V}$. The channel width is $\rm{Wch}=4.3\rm{nm}$, $\rm{Wch}=7.1\rm{nm}$, and $\rm{Wch}=9.8\rm{nm}$, from the first to the third row. The channel length is Lch=Wch, Lch=2Wch, Lch=3Wch, and Lch=4Wch, from the first to the fourth column.}
\label{IV_curve_all}
\end{figure*}

With the MS basis, the transistor I-V curves are computed in the Poisson-MS-NEGF approach. 
The channel length ($L_{\rm{ch}}$) is scaled as $1\times$, $2\times$, $3\times$, and $4\times$ of the channel width ($W_{\rm{ch}}$), covering short-channel to long-channel situations. The Si MOSFETs have source and drain doping density $1\times10^{20}\rm{cm}^{-3}$ while the InGaAs MOSFETs have slightly lighter doping density $5\times10^{19}\rm{cm}^{-3}$. The EOT is 1nm for all cases. The transfer characteristics of all cases are plotted in Fig.~\ref{IV_curve_all}. The current is normalized by $W_{\rm{ch}}$. Other current normalization methods can be used but will not affect our comparison. We focus on target OFF current level $I_{\rm{OFF}}=0.1\mu\rm{A}/\mu\rm{m}$ and supply voltage $V_{\rm{DD}}=0.5\rm{V}$.

For small cross section (Fig.~\ref{IV_curve_all}, first row), it is observed that at long channel lengths Si has larger ON current ($I_{\rm{ON}}$) than InGaAs, although their subthreshold swing (SS) are very similar (approach 60mV/dec room-temperature limit). The larger $I_{\rm{ON}}$ of Si can be attributed to its larger DOS compared with InGaAs. In fact, at such small cross section, the InGaAs nanowire MOSFET operates at the quantum capacitance limit~\cite{kim2015comprehensive}. At short channel lengths, the SS of InGaAs degrades faster than Si and therefore Si still has a larger $I_{\rm{ON}}$. It is known that the degradation of SS at short channel lengths is due to the reduced gate control as well as increased SDT. The InGaAs has smaller electron effective mass and thus severer SDT than Si. For medium cross section (Fig.~\ref{IV_curve_all}, second row), it is observed that $I_{\rm{ON}}$ of Si is larger than InGaAs at both long channel and short channel situations, similar to the small cross section case. However, the difference between Si and InGaAs is smaller compared with the small cross section case. For large cross section (Fig.~\ref{IV_curve_all}, third row), at long channel length there is hardly any difference observed between Si and InGaAs. At short channel lengths, Si again possesses SS and $I_{\rm{ON}}$ advantages over InGaAs.

\begin{figure}[]
\centering
\includegraphics[width=3.31in]{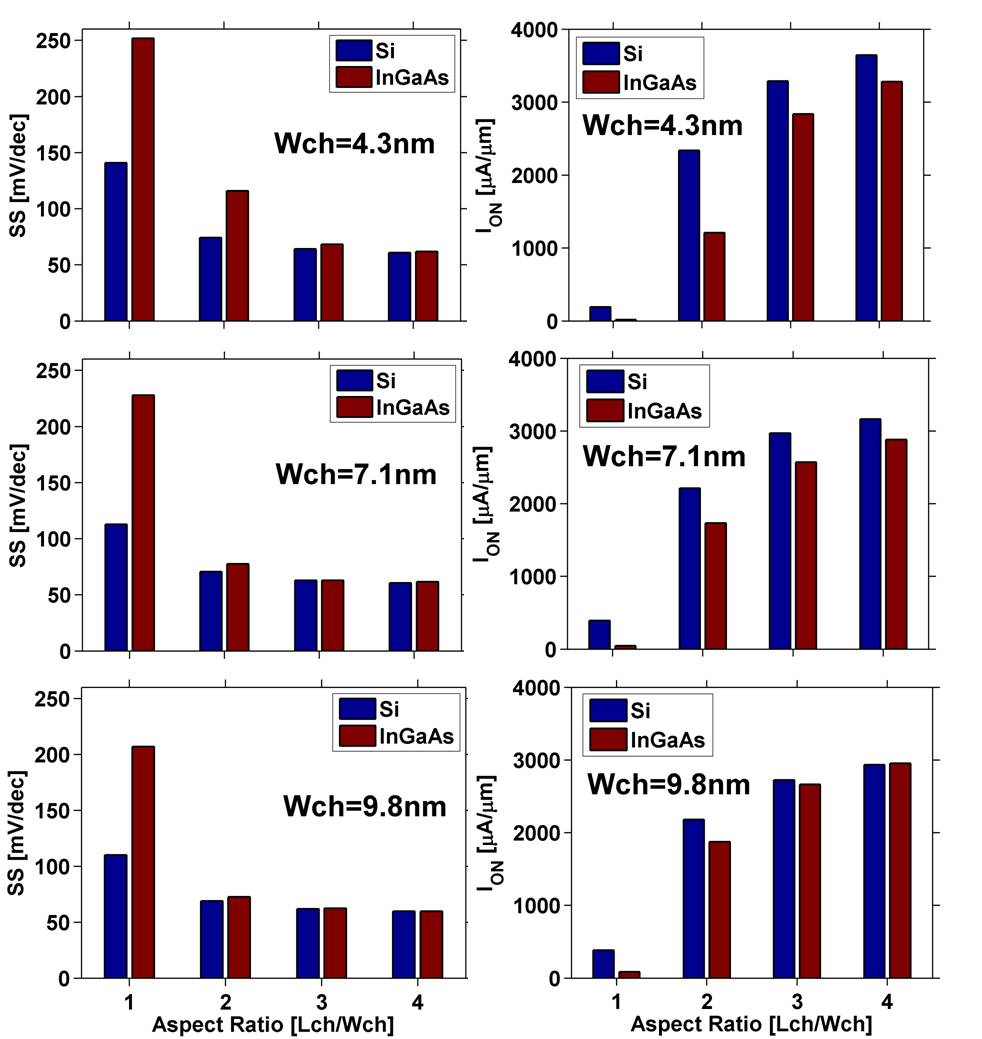}
\caption{Subthreshold swing (SS) and ON current ($I_{\rm{ON}}$) of the Si and InGaAs nanowire nMOSFETs at $V_{\rm{DD}}=0.5\rm{V}$ and $I_{\rm{OFF}}=0.1\mu\rm{A}/\mu\rm{m}$. Top: $\rm{Wch}=4.3\rm{nm}$, middle: $\rm{Wch}=7.1\rm{nm}$, bottom: $\rm{Wch}=9.8\rm{nm}$.}
\label{SS_Ion_all}
\end{figure}

The SS and $I_{\rm{ON}}$ are further summarized in Fig.~\ref{SS_Ion_all}. It can be concluded that (1) Si has smaller SS, especially at short channel lengths; (2) Si has larger $I_{\rm{ON}}$, except that at large cross section and long channel length InGaAs shows slightly larger $I_{\rm{ON}}$ than Si; (3) Aspect ratio ($L_{\rm{ch}}/W_{\rm{ch}}$) of 3 is critical for high performance, above which the performances start to saturate. Different channel orientation and/or confinement orientations may change the band structures and thus the quantitative results~\cite{neophytou2008bandstructure,kim2015comprehensive}, but the general trend here should remain the same. It should also be mentioned that source/drain doping density has a large impact on the depletion length of III-V nMOSFETs, thus it needs to be optimized for reducing the SDT, important at ultra-short channel lengths~\cite{kim2015source}.

\begin{figure}[]
\centering
\includegraphics[width=3.31in]{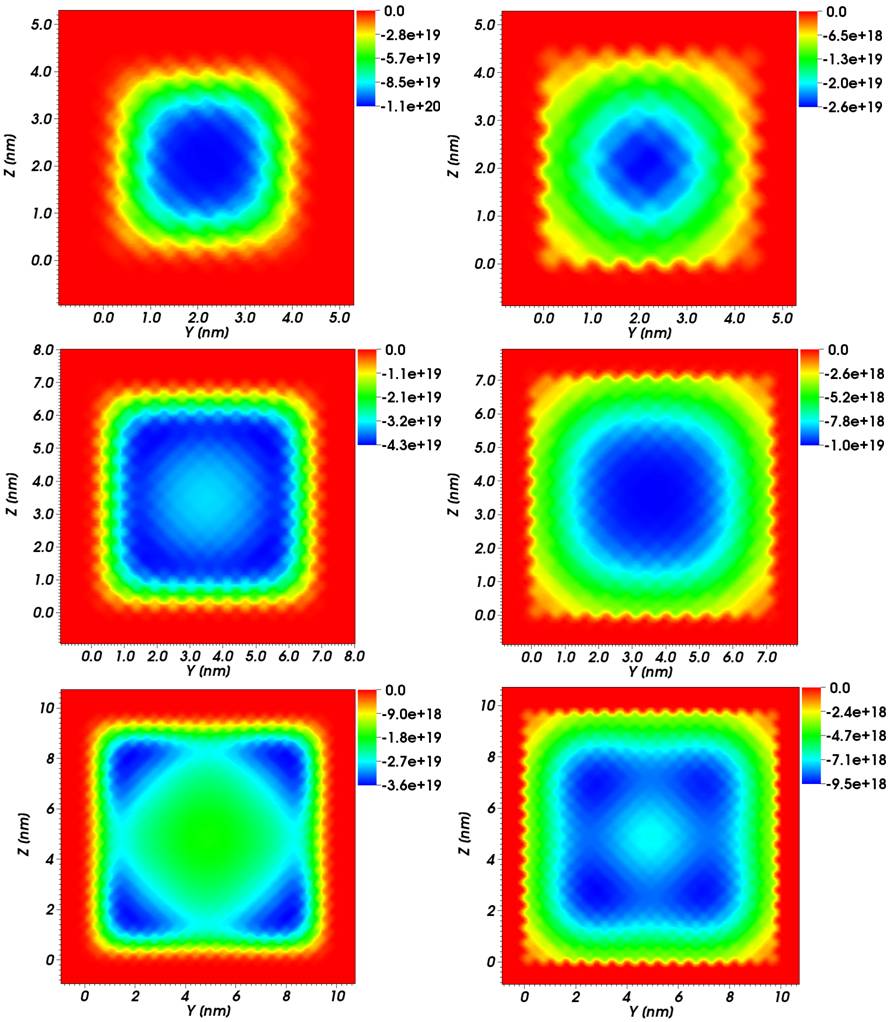}
\caption{Charge density distribution at top of the barrier (ON state, long channel). Left: Si, right: InGaAs. Top: $4.3\rm{nm}\times4.3\rm{nm}$ cross section, middle: $7.1\rm{nm}\times7.1\rm{nm}$ cross section, bottom: $9.8\rm{nm}\times9.8\rm{nm}$ cross section.}
\label{charge_density_all}
\end{figure}

It is interesting to visualize the charge density distribution at the top of the barrier for various cross sections (Fig.~\ref{charge_density_all}). It is observed that, (1) for both Si and InGaAs, as the cross section becomes larger, the charge starts to accumulate toward the corners; (2) the charge of Si distributes closer to the corners than InGaAs; (3) Si has a few times larger change density than InGaAs. Such observations are consistent with the facts that Si has larger density of states and heavier effective mass than InGaAs.

Finally, we note that the largest device in this study, {\it i.e.}, the $9.8\rm{nm}\times9.8\rm{nm}$ cross section Si MOSFET with 39.1nm channel length (54.3nm tolal device length), consists of about 259,200 Si atoms. The basis optimization took 0.61 hours and the I-V curve simulation (8 bias points) took 22.6 hours, both using the same 6 nodes cluster, 4 MPI processes per node, and 4 cores per MPI process.

\section{Conclusions}
A robust and parallel algorithm based on the Mil'nikov approach is developed to optimize the mode space basis for atomistic nanowires. With this algorithm, reliable mode space tight binding basis can be generated efficiently for nanowires with cross section up to $10\rm{nm}\times10\rm{nm}$. Basis reduction ratio of a few percent has been achieved and NEGF simulation with speed up factor of over 300 has been demonstrated. This enables accurate full-band quantum transport simulation of realistically-sized nanowire transistors in a small computer cluster. As an application of this method, ballistic I-V curves of InGaAs and Si nanowire MOSFETs are compared for a wide range of device dimensions, it is found that InGaAs MOSFETs outperform Si MOSFETs only when the cross section is above $10\rm{nm}\times10\rm{nm}$ and the channel length is greater than 40nm. A more accurate performance comparsion requires scattering effects to be taken into account, which is feasible in the mode space NEGF framework, a topic for future study.

% If in two-column mode, this environment will change to single-column format so that long equations can be displayed. 
%\begin{widetext}
%$$\mbox{put long equation here}$$
%\end{widetext}

% Figures should be put into the text as floats. 
% Use the graphics or graphicx packages (distributed with LaTeX2e).
% Put the label that you will use with \ref{} command in the braces of the \label{} command.
%
% \begin{figure}
% \end{figure}

% \begin{table}
% \end{tabular}
% \end{table}

\begin{acknowledgments}
This work uses nanoHUB.org computational resources operated by the Network for Computational Nanotechnology funded by the U.S. National Science Foundation under Grant EEC-0228390, Grant EEC-1227110, Grant EEC-0634750, Grant OCI-0438246, Grant OCI-0832623, and Grant OCI-0721680. NEMO5 developments were critically supported by an NSF Peta-Apps award OCI-0749140 and by Intel Corp. J. Z. Huang acknowledges support of NSFC Grant No.61431014 of China via Zhejiang University. M. Povolotskyi acknowledges support of NSF (grants 1639958, 1509394) and SRC (grant 2694.003). J. Z. Huang thanks D. Lemus for help on analyzing the timing of the basis optimization, and thanks B. Novakovic for helpful discussions.
\end{acknowledgments}

\nocite{*}
\bibliography{Robust_Mode_Space_Approach}

\end{document}